\documentclass[runningheads,a4paper]{llncs}

\usepackage{graphicx}
\usepackage{pstricks}

\usepackage{hyphenat}

\usepackage{hyperref}
 \hypersetup{colorlinks=false}
\usepackage{graphicx}
\newcommand{\ignore}[1]{}

\newcommand{\Poly}{\mathrm{P}}
\newcommand{\NP}{\mathrm{NP}}
\newcommand{\APX}{\mathrm{APX}}
\renewcommand{\epsilon}{\varepsilon}
\usepackage{amssymb}
\usepackage{amsmath}
\usepackage{graphicx}
\usepackage{verbatim} 
\usepackage{url}
\usepackage[lined,algonl,boxed]{algorithm2e}
\usepackage{hyperref}
 \hypersetup{colorlinks=false}
\usepackage{graphicx}

\usepackage{amssymb}
\usepackage{graphicx}

\usepackage{url}
\urldef{\mailsa}\path|{sroy7}@buffalo.edu|

\usepackage{hyperref}
 \hypersetup{colorlinks=false}
\usepackage{graphicx}
\usepackage{pstricks}
\usepackage{amssymb}
\usepackage{graphicx}
\usepackage{verbatim} 
\usepackage{url}
\usepackage[lined,algonl,boxed]{algorithm2e}
\usepackage{hyperref}
 \hypersetup{colorlinks=false}
\usepackage{graphicx}
\begin{document}

\mainmatter  % start of an individual contribution

\title{On Approximability of Block Sorting}

\titlerunning{On Approximability of Block Sorting}

%\author{Swapnoneel Roy%
%\thanks{Research supported in part by NSF grant CCF-0844796.}%
%}
%
%\authorrunning{Swapnoneel Roy}


%\institute{Department of Computer Science and Engineering,\\
%University at Buffalo, The State University of New York,\\
%Buffalo, New York 14260.\\
%\url{sroy7@buffalo.edu}}

\author{N. S. Narayanaswamy\inst{1} \and Swapnoneel Roy\inst{2}%
\thanks{Research supported in part by NSF grant CCF-0844796. A part of this work was done when this author was with Dept of CS$\&$E, IIT Madras.}%
}

\authorrunning{Narayanaswamy et al.}

%\institute{Department of Computer Science and Engineering, Indian Institute of Technology Madras, Chennai, Tamil Nadu, India 600036. \url{swamy@iitm.ac.in}.
%\and 
%Department of Computer Science and Engineering, University at Buffalo, The State University of New York, Buffalo, New York 14260. \url{sroy7@buffalo.edu}.

\institute{Department of Computer Science and Engineering,\\
Indian Institute of Technology Madras, Chennai, TN 600036, India.
\and
Department of Computer Science and Engineering, \\
University at Buffalo, The State University of New York, Buffalo, NY 14260, USA.
}


\toctitle{On Approximability of Block Sorting}
\tocauthor{Contents}
\maketitle
% or an extended abstract of the paper has appeared elsewhere}
%%%%%%%%%%%%%%%%%%%%%%%%%%%%%%%%%%%%%%%%%%%%%%%%%%%%%%%%%%%%%%%%%%%%%%%%%%%

%% the abstract has to PRECEDE the command \maketitle:
%% be sure not to issue the \maketitle command twice!

\begin{abstract} {\sc Block Sorting} is a well studied problem, motivated by its applications in Optical Character Recognition (OCR), and Computational Biology. {\sc Block Sorting} has been shown to be $\NP$-Hard, and two separate polynomial time $2$-approximation algorithms have been designed for the problem. But questions like whether a better approximation algorithm can be designed, and whether the problem is $\APX$-Hard have been open for quite a while now.

\noindent
In this work we answer the latter question by proving {\sc Block Sorting} to be Max-$\mathcal{SNP}$-Hard ($\APX$-Hard). The $\APX$-Hardness result is based on a linear reduction of {\sc Max-$3$SAT} to {\sc Block Sorting}. We also provide a new lower bound for the problem via a new parametrized problem {\sc $k$-Block Merging}.  
%\keywords{
%Sorting by $1$-Block Interchanges \and strip exchanging \and Block Interchanges \and combinatorial optimization \and sorting by strip moves \and block sorting \and NP-Completeness \and optical character recognition \and genome rearrangement}
\end{abstract}

\section{Introduction}\label{intro1}

The {\sc Block Sorting} problem is a combinatorial optimization problem to find out the minimum number of block moves required to sort a given permutation $\pi$. A block is a maximal substring of $\pi$, which is also a substring of the sorted (identity) permutation $id$. {\sc Block Sorting} is motivated by its applications in optical character recognition~\cite{paper1}$\mbox{-}$\cite{blocksort05}. In optical character recognition, text regions referred to as {\em zones} are identified. The ordering of the zones is very important. But in practice, the output generated by any zoning algorithm is frequently different from the correct order. To measure how good a zoning algorithm is, we need to find the minimum number of steps required to transform the string generated by the zoning algorithm, to the correct string. The problem of obtaining the number of steps to convert the given string into the correct string is equivalent to {\sc Block Sorting.} Hence it is very important to design efficient algorithms for {\sc Block Sorting}, and know more about its computational complexity. 

\noindent
{\sc Block Sorting} is also gains much importance from the fact that it is a nontrivial variation of a very well known problem {\sc Sorting by Transpositions} which is motivated by the study of genome rearrangements in computational biology. In transpositions, we are allowed to move any substring of $\pi$ to a different position at each step~\cite{paper2}. {\sc Sorting by Transpositions} optimizes the number of such moves to sort $\pi$. It is easy to see that a block move is a transposition, but not the vice versa. {\sc Sorting by Transpositions} has been recently shown to be $\NP$-Hard~\cite{paper4}. The best known algorithm for {\sc Sorting by Transpositions} has an approximation ratio of $1.375$~\cite{paper3}. It is not known yet whether {\sc Block Sorting} approximates {\sc Sorting by Transpositions} to any factor better than $3$. But it is known that optimal transpositions never need to {\em break} existing blocks~\cite{paper5}. This shows how the two problems are closely related. The study of the computational complexity of {\sc Block Sorting} therefore might provide us with more insight into the complexity of {\sc Sorting by Transpositions}. It is still not known whether {\sc Sorting by Transpositions} is $\APX$-Hard\footnote{Defined in Appendix~\ref{glossary}}, or it admits a PTAS\footnotemark[\value{footnote}].

\noindent
{\sc Block Sorting} is also closely related to another problem called {\sc Sorting by Short Block-Moves}. In a short block move, we are allowed to move an element of $\pi$ to at most two positions away from its original position. {\sc Sorting by Short Block-Moves} optimizes the number of such moves required to sort $\pi$~\cite{paper6}. The problem is motivated by its applications in the study of genome rearrangements and in the design of interconnection networks. It is easily observed that a short-block move is also a block move, but not the vice-versa. The complexity of {\sc Sorting by Short Block-Moves} is still open. It has been studied extensively, and recently a PTAS has been designed for the problem~\cite{paper7}. We believe that our results on the complexity of {\sc Block Sorting} could help resolve the computation complexity of {\sc Sorting by Short Block-Moves}, which has been open for close to a decade and a half now.

\section{Overview of the results and techniques}
\noindent
The set $\{1,2,\cdots,n\}$ is denoted by $[n]$, and let $S_n$ denote the set of all permutations over $[n]$, and $id_n$ the sorted or identity permutation of length $n$. The given permutation $\pi \in S_n$ to be sorted is represented as a string $\pi_1\pi_2\cdots\pi_n$ without loss of generality.

\begin{definition}[Block]\label{blocks}
A block is a maximal substring of a given permutation $\pi$, which is also a substring of the identity permutation $id$. 
\end{definition}
\noindent
As an example, the permutation 8 2 5 6 3 9 1 4 7 contains $8$ blocks, and $\fbox{5 6}$ is the only block of length more than one. A block move picks up a block and places it elsewhere in the permutation. A block sorting schedule is a sequence of block moves to sort a given permutation $\pi$. The minimum number of such moves required is called the block sorting distance $bs(\pi)$ of permutation $\pi$. An example of the block move of moving block $1$ to block $2$ for \\$\pi=$ 3 1 4 6 2 5 7 9 8 10 is shown in Figure\ref{ex1} in Appendix~\ref{sortingex}.

\noindent
Any block in $\pi$ could be replaced by a single element without loss of generality. Hence the permutation 7 2 5 3 8 1 4 6 is equivalent to  8 2 5 6 3 9 1 4 7. We can do this because we do not break blocks once they get joined to form larger blocks in a block move. The reduced permutation is termed as {\em reduced permutation}~\cite{blocksort03} or a {\em kernel permutation}~\cite{blocksort06}. 

\noindent
\textsc{Block Sorting} can be stated as:
\begin{center}
\framebox{
\begin{tabular}{l}
\noindent {\sc Block Sorting Problem}
\\ \noindent {\sc Input:} A permutation $\pi$ and an integer $m$.
\\ \noindent {\sc Question:}  Is $bs(\pi) \leq m$?
\end{tabular}
}
\end{center}
\noindent
In~\cite{blocksort06}, it has been formally proved that block-sorting $\pi$ is equivalent to block-sorting its kernel $ker(\pi)$. That is $bs(\pi)=bs(ker(\pi))$. Also it was shown in~\cite{blocksort06}, that in an optimal block-sorting sequence, we never need to break apart an existing block at any step. That is the block-sorting distance remains the same, even if we allow block-sorting moves which do not necessarily join blocks, or which breaks any previously joined blocks.

\noindent 
{\sc Block Sorting} was proved to be $\NP$-Hard via a reduction from $3SAT$ in~\cite{blocksort03}. We reduce {\sc Max-$3$SAT} to {\sc Block Sorting} via a linear reduction and prove it to be Max-$\mathcal{SNP}$-Hard. We achieve this by proving a new technical lemma (Lemma~\ref{disconn}) for block sorting. We prove that the number of moves in any block sorting schedule for any permutation is at least the sum of the number of reversals in the permutation, and the number of disconnected components in the red-blue graph constructed from that schedule. Our results show {\sc Block Sorting} does not admit a PTAS unless $\Poly = \NP$.

\begin{definition}[Reversal] 
In a permutation $\pi$, a reversal is a pair of consecutive elements $a$$b$ such that $a>b$. Formally $a$ and $b$ form a reversal in $\pi$ if $a>b$ and $\pi_b = \pi_a+1$.
\end{definition}
\noindent
Let the number of reversals in $\pi$ be $rev(\pi)$. In~\cite{blocksort03}, it has been shown that a block sorting sequence of length $rev(\pi)$ is optimal, since the block sorting distance $bs(\pi) \geq rev(\pi)$. 

\noindent
In~\cite{blocksort03} the authors had constructed permutation $\pi$ from an arbitrary $3SAT$ boolean formula $\Phi$, such that

\noindent
$\bullet$ $\Phi$ is satisfiable if and only if $bs(\pi)=rev(\pi)$.

\noindent
In this work we construct permutation $\pi$ from an arbitrary {\sc Max-3SAT} instance, a boolean formula $\Phi$ with $m$ clauses, such that

\noindent
$\bullet$ If all the $m$ clauses of $\Phi$ are satisfiable, then $bs(\pi)=rev(\pi)$.

\noindent
$\bullet$ If at most $m-c$ clauses of $\Phi$ are satisfiable, then $bs(\pi) \geq rev(\pi)+c$.

\noindent
The above proves \textsc{Block Sorting} to be Max-$\mathcal{SNP}$-Hard, which is one of our results. 

\noindent
The question whether {\sc Block Sorting} is $\APX$-Hard was open for quite sometime now, and we answer it in this work. We believe, this result will also provide us more insight on the complexity of the more general {\sc Sorting by Transpositions}, which has been recently proved to be $\NP$-Hard. Also we might use a similar reduction technique to prove the hardness of {\sc Sorting by Short-Block Moves}, whose complexity remains unresolved till date.  

\noindent 
The problem {\sc Block Merging} has been introduced in~\cite{blocksort06} in conjunction to obtaining a factor $2$ approximation algorithm for {\sc Block Sorting}. The permutation $\pi$ can be uniquely decomposed into maximal increasing subsequences. The input to {\sc Block Merging} is the set of these increasing subsequences $\mathbb{S}_{\pi}$. If $\pi=$ 8 2 5 6 3 9 1 4 7, then $\mathbb{S}_{\pi}=\{(8), (2,5,6), (3,9), (1,4,7)\}$ is an input instance for {\sc Block Merging}. The goal of {\sc Block Merging} is to transform $\mathbb{S}_{\pi}$ to the multiset $\mathbb{M}_{n}=\{id_n, \epsilon,\cdots,\epsilon\}$ using the minimum number of block moves. A block move on a block is permitted, if and only if the block is contained in {\em at most one} increasing subsequence. At the beginning, every block in $\mathbb{S}_{\pi}$ is contained in exactly one increasing subsequence. But during the execution of a block-merging schedule, a block might get {\em fragmented} over several increasing subsequences. Let $bm(\mathbb{S}_{\pi})$, called the block-merging distance of $\pi$, be the minimum such moves to transform $\mathbb{S}_{\pi}$ to $\mathbb{M}_{n}$. 

\noindent
In~\cite{blocksort06} the authors proved a new lower bound for {\sc Block Sorting} via {\sc Block Merging} as

\noindent
$\bullet$ $bs(\pi) \geq \frac{bm(\mathbb{S}_{\pi})}{2}$.

\noindent
They further proved that {\sc Block Merging} $\in \Poly$ to design a $2$-approximation algorithm for {\sc Block Sorting} via {\sc Block Merging}. 

\noindent
We parametrize {\sc Block Merging} to formulate another problem {\sc $k$-Block Merging}, for any integer $k$. In {\sc $k$-Block Merging}, we have the same input and goal as {\sc Block Merging}. But we allow a block to be moved if it is contained in {\em at most $k$} increasing subsequences. Hence {\sc Block Merging} is {\sc $k$-Block Merging} with $k=1$. Let $k$-$bm(\mathbb{S}_{\pi})$ be the $k$-block merging distance for $\pi$. 

\noindent
We prove a new lower bound for {\sc Block Sorting} via {\sc $k$-Block Merging} as

\noindent
$\bullet$ $bs(\pi) \geq \frac{k\mbox{-}bm(\mathbb{S}_{\pi})}{1+\frac{1}{k}}$.

\noindent
In other words, {\sc $k$-Block Merging} approximates {\sc Block Sorting} by a factor of $1+\frac{1}{k}$. We know {\sc $k$-Block Merging} $\in \Poly$ for $k=1$. But we do not know anything about its complexity for $k>1$. If we can prove it to be polynomial for $k=2$, we have a $1.5$-approximation algorithm for {\sc Block Sorting}. To recall, the best known approximation algorithms for {\sc Block Sorting} have a factor of $2$. On the other hand if we prove it to be $\NP$-Hard for $k=2$, then we actually prove {\sc Block Sorting} to be in-approximable to within a factor of $1.5$, which improves the integrality gap we achieve in our reduction.  

\section{The Red-Blue Graph for \textsc{Block Sorting}}\label{redblue}
\noindent
Given $\pi$, and a block sorting sequence $\mathcal{S}$ for $\pi$, we construct the red-blue graph $G(\pi, \mathcal{S})$, where the vertices are the blocks of $\pi$ in the following way:
\noindent
\begin{enumerate}
\item A {\bf blue edge} is constructed between the participating blocks $a$ and $b$ of each reversal $(a,b)$.
\item A {\bf red edge} is  constructed between two blocks $a$ and $b$ if:
\begin{itemize}
\item $a<b$ and $\pi_a<\pi_b$, 
\item $a$ and $b$ are joined in the sequence $\mathcal{S}$ before either is moved, and
\item if $\pi_a<\pi_c<\pi_b$, then block $c$ is moved before $a$ and $b$ are joined.
\end{itemize}
\end{enumerate}
\noindent
The intuition behind the red edges is to treat the two blocks participating as already in their correct positions~\cite{blocksort03}. We need to move only the other blocks in the block sorting schedule. Thus we effectively {\em save} one move per red edge. Hence the construction of $G(\pi, \mathcal{S})$ is dependent on the given block sorting schedule $\mathcal{S}$ on $\pi$. 
\noindent
Now we present a few properties about the red edges proved in ~\cite{blocksort03}. We will state them here without proof. $\pi(i)$ is the position of block $i$ in $\pi$.

\begin{lemma}~\cite{blocksort03}\label{acyclic} 
For any $\pi$, $G(\pi)$ is acyclic over both red and blue edges.
\end{lemma}

\begin{lemma}~\cite{blocksort03}\label{red1}
Any node $x$ can have a red degree of at most $2$. One from $y$ to $x$ where $y < x$ another from $x$ to $z$ where $x < z$.
\end{lemma}

\begin{lemma}~\cite{blocksort03}\label{red2}
If $\pi(a)<\pi(c)<\pi(b)<\pi(d)$, there cannot be both a red edge from $a$ to $b$, and a red edge from $c$ to $d$.
\end{lemma}

\begin{lemma}~\cite{blocksort03}\label{red3}
If $a<c<b<d$, there cannot be both a red edge from $a$ to $b$, and a red edge from $c$ to $d$.
\end{lemma}

\begin{lemma}~\cite{blocksort03}\label{red4}
If $a<c<d<b$ and $\pi(c)<\pi(a)<\pi(b)<\pi(d)$, there cannot be both a red edge from $a$ to $b$, and a red edge from $c$ to $d$.
\end{lemma}

\noindent
A pair of red edges (or pairs of elements $(a,b)$ and $(c,d)$ in $\pi$) which violate any of Lemmas~\ref{red1}, ~\ref{red2}, ~\ref{red3}, or ~\ref{red4}, {\em cross} each other.

\noindent
\begin{definition}[Perfect block sorting]~\cite{blocksort03}
A perfect block sorting schedule $\mathcal{S}_{perfect}$ on $\pi$ is a block sorting schedule which sorts $\pi$ in $rev(\pi)$ moves. Note that $rev(\pi)$ is equal to the number of blue edges in graph $G(\pi, \mathcal{S}_{perfect})$. 
\end{definition}

\noindent
Perfect block sorting is optimal since $bs(\pi) \geq rev(\pi)$.
\begin{lemma}\label{opt}~\cite{blocksort03}
There exists a perfect block sorting schedule $\mathcal{S}_{perfect}$ on $\pi$ if and only if $G(\pi, \mathcal{S}_{perfect})$ is a tree.
\end{lemma}

\noindent
Given $\pi$, the blue edges of $\pi$ are always fixed, in $G(\pi, \mathcal{S})$, for any block sorting schedule $\mathcal{S}$. Hence when we refer to blue edges, we would talk about only $\pi$ instead of $G(\pi, \mathcal{S})$. The red edges will vary for different block sorting schedules. 
\begin{definition}[Blue component]
In permutation $\pi$, all the blue edges connect elements which are adjacent to each other. We define the components connected by zero or more blue edges as blue components of a red-blue graph. All the blue components are substrings of $\pi$.
\end{definition}

\noindent
As an example, for $\pi=$ 8 2 5 6 3 9 1 4 7, the blue components are $\{8$ $2\}$, $\{\fbox{5 6}$ $3\}$, $\{9$ $1\}$, $\{4\}$, and $\{7\}$. The number of blue components in $\pi$ is equal to the difference between the number of blocks in $\pi$ and $rev(\pi)$. Formally, $\#blue\mbox{-}components(\pi) = \#blocks(\pi)-rev(\pi)$. In a red-blue graph of a perfect block sorting schedule, all these blue components are connected, since the graph is a tree. Hence there can be at most $\#blocks(\pi)-rev(\pi)-1$ red edges in any red-blue graph of $\pi$. Intuitively, we see that the more are the number of disconnected blue components in the red-blue graph $G(\pi, \mathcal{S})$, the more are the number of moves in $\mathcal{S}$. Each red edge saves one move. Let the number of disconnected components in any graph be defined as $\#disconnected\mbox{-}components(G(\pi, \mathcal{S})) = \#connected\mbox{-}components(G(\pi, \mathcal{S})) - 1$. $d$ disconnected components signifies an absence of $d$ red edges to connect them to the rest of the graph. Therefore $d$ disconnected component signifies at least $d$ more moves for $\mathcal{S}$ than $\mathcal{S}_{perfect}$. We formally prove this in Lemma~\ref{disconn}. The length of $\mathcal{S}$ is the number of moves in $\mathcal{S}$.

\begin{lemma}\label{disconn}
Length of any block sorting schedule $\mathcal{S} \geq rev(\pi) + \#disconnected\mbox{-}components(G(\pi, \mathcal{S}))$ , for any block sorting schedule $\mathcal{S}$ on $\pi$.
\end{lemma}

\noindent
The proof of Lemma~\ref{disconn} is in Appendix~\ref{sec2}. Theorem~\ref{tighterlb} follows from Lemma~\ref{disconn}.

\begin{theorem}\label{tighterlb}
For any permutation $\pi$, if $\#disconnected\mbox{-}components(G(\pi, \mathcal{S})) \geq d$ for every possible block sorting schedule $\mathcal{S}$ on $\pi$, then $bs(\pi) \geq rev(\pi)+d$ for that permutation $\pi$.
\end{theorem}
\noindent
Theorem~\ref{tighterlb} says that for any $\pi$ if we show that the number of disconnected blue components in the red-blue graph of any block sorting schedule is at least $d$, then the lower bound for block sorting for $\pi$ is at least $d$ more than $rev(\pi)$. We are now in a position to prove the $\APX$-hardness of {\sc Block Sorting}.

\section{\textsc{Block Sorting} is Max-$\mathcal{SNP}$-Hard}\label{npc}
\noindent
We use the construction from~\cite{blocksort03} to reduce {\sc Max-$3$SAT} to {\sc Block Sorting}. Consider an instance of {\sc Max-$3$SAT} consisting of a boolean formula $\Phi = \mathcal{C}^1 \mathcal{C}^2 \cdots \mathcal{C}^m$ of $n$ variables and $m$ clauses $\mathcal{C}^1, \mathcal{C}^2, \cdots, \mathcal{C}^m$. A permutation $\pi$ of $8m+4n+1$ elements was constructed from $\Phi$ by introducing an ordered alphabet $\Sigma_{n,m}$ with $4nm+2m+4n+1$ elements in~\cite{blocksort03}. A block sorting schedule $\mathcal{S}$ of length $6m+2n-1$ has been shown to exist for $\pi$ if and only if $\Phi$ was satisfiable.

\noindent
Here we use that construction to show the following:

\begin{enumerate}

\item Max-$3SAT(\Phi)=m$ $\Longrightarrow$ $bs(\pi)=6m+2n-1$.

\item Max-$3SAT(\Phi) \leq m-c$ $\Longrightarrow$ $bs(\pi) \geq 6m+2n-1+c$.

\end{enumerate}

\noindent
The alphabet $\Sigma_{n,m}$ consists of the following elements:

\begin{enumerate}
\item {\bf Term symbols:} $p_i^j, \bar{p}_i^j, q_i^j, \bar{q}_i^j$, $\forall$ $1 \leq i \leq n$, and $1 \leq j \leq m$. $p_i^j, \bar{p}_i^j$ and $q_i^j, \bar{q}_i^j$ are called left and right term symbols respectively.

\item {\bf Clause control symbols:} $\ell^j$ and $r^j$ $\forall$ $1 \leq j \leq m+n$.

\item {\bf Variable control symbols:} $u_i$ and $\upsilon_i$ $\forall$ $1 \leq i \leq n$.

\item {\bf Separator symbol:} $s$.
\end{enumerate}

\noindent
The ordering on $\Sigma_{n,m}$ is generated by the following rules:

\begin{enumerate}
\item $u_i < p_i^k < p_i^j < \bar{p}_i^k < \bar{p}_i^j  < q_i^j < q_i^k < \bar{q}_i^j < \bar{q}_i^k < \upsilon_i < s$ $\forall$ $1 \leq i \leq n$, and $1 \leq j < k \leq m$.

\item $\upsilon_{i-1} < u_i$ $\forall$ $1 \leq i \leq n$.

\item $s< \ell^k < r^k < \ell^j < r^j$ $\forall$ $1 \leq j < k \leq m+n$.
\end{enumerate}

\noindent
Let the names of each variable be $x_i$ $\forall$ $1 \leq i \leq n$. Each clause is assumed to be of the form $(z_a \vee z_b \vee z_c)$, $a>b>c$ without loss of generality, where $z_i$ is either $x_i$ or $\bar{x}_i$. The simple encoding of a clause uses symbols $p_i$  and $q_i$ and $\bar{p}_i$  and, $\bar{q}_i$ respectively for literal $x_i$ and, $\bar{x}_i$. It consists of eight symbols starting with an $\ell$, and ending with $r$. The remaining symbols are the terms symbols, that start with $p$'s and end with the $q$'s. As an example the simple encoding of the clause $(x_5 \vee x_3 \vee \bar{x}_2)$ is $\ell p_5 p_3 \bar{p}_2 q_5 q_3 \bar{q}_2 r$. For the real encoding of a clause, the index of that clause is inserted as its superscript. If $\mathcal{C}^4 = (x_5 \vee x_3 \vee \bar{x}_2)$, then its encoding would be $\ell^4 p_5^4 p_3^4 \bar{p}_2^4 q_5^4 q_3^4 \bar{q}_2^4 r^4$. The clause encodings are in the order $\mathcal{C}^1$ followed by $\mathcal{C}^2$ till $\mathcal{C}^m$ in $\pi$.

\noindent
After the real encoding of each clauses, the control sequences $\ell^{m+i}u_i\upsilon_ir^{m+i}$ are added $\forall$ $1 \leq i \leq n$. Finally the element $s$ is added as the first element. Hence the derived permutation $\pi$ contains $s$, followed by the encodings of the clauses in order, followed by the control sequences $\forall$ $i$ from $1$ to $n$. An example of the reduction of $\Phi = (\bar{x}_3 \vee x_2 \vee x_1)\wedge(x_4 \vee \bar{x}_3 \vee \bar{x}_2)$ is shown in Figure~\ref{reduction} of Appendix~\ref{exred}. In this figure, we have two clause components in $\pi$, for the two clauses in $\Phi$.

\begin{lemma}\label{num-blue-edge}~\cite{blocksort03}
$G(\pi, \mathcal{S})$ has $6m+2n-1$ blue edges for any block sorting schedule $\mathcal{S}$ on $\pi$, and hence $bs(\pi) \geq 6m+2n-1$.
\end{lemma} 

\noindent
Lemma~\ref{num-blue-edge} follows from the fact that $rev(\pi) = 6m+2n-1$.  We state and prove lemmas in this section in a {\em top-down} manner. Lemma~\ref{unsat}, Lemma~\ref{unsatdisconn}, and Lemma~\ref{left} are stated and proved first. Next we state and prove the lemmas which lead to Lemma~\ref{unsat}, Lemma~\ref{unsatdisconn}. The omitted proofs appear in Appendix~\ref{sec3}. 

\noindent
The blue components of any $G(\pi, \mathcal{S})$ are:
\begin{enumerate}
\item The blue component with the $s$ symbol. There is $1$ such component.
\item The blue components with the $p_i^j$ (or $\bar{p}_i^j$) symbols, for each $1 \leq j \leq m$, $1\leq i \leq n$. There are $m$ such components.
\item The blue components with the $q_i^j$ (or $\bar{q}_i^j$) symbols, for each $1 \leq j \leq m$, $1\leq i \leq n$. There are $m$ such components.
\item The blue components with the $u^i$  symbols, for each $1\leq i \leq n$. There are $n$ such components.
\item The blue components with the $\upsilon^i$  symbols, for each $1\leq i \leq n$. There are $n$ such components.
\item The blue component with the $r^{m+n}$ symbol. There is $1$ such component.
\end{enumerate}
\noindent
Hence there are in total $2m+2n+2$ blue components of any $G(\pi, S)$ for $\pi$. The red-blue graph of a perfect schedule on $\pi$ is a tree. Since $G(\pi, \mathcal{S})$ is acyclic~\cite{blocksort03}, there can be at most $2m+2n+1$ red edges to connect these $2m+2n+2$ blue components. 

\noindent
The three $q_i^j$ symbols $1\leq i \leq n$, for the components of each clause $\mathcal{C}^j$ for $1 \leq j \leq m$, are joined by two blue edges, and form a blue component in any red blue graph $G(\pi, \mathcal{S})$. We call these blue components containing $q$ symbols for each clause component. Therefore out of the $2m+2n+2$ blue components, $m$ are blue components containing $q$ symbols. 

\begin{lemma}\label{one-on-one}
There is a one-on-one correspondence between all possible assignments in $\Phi$ to all possible arrangements of the set of red edges $\{(p_i^j, q_i^j)$ and $(\bar{p}_i^j, \bar{q}_i^j), 1\leq i \leq n, 1\leq j \leq m\}$.
\end{lemma}

\begin{lemma}\label{unsat}
Given a satisfying assignment of $\Phi$, if a clause $\mathcal{C}^j \in \Phi$ is unsatisfied, we cannot have any red edges between any $(p_i^j, q_i^j)$ (or $(\bar{p}_i^j, \bar{q}_i^j)$), $1\leq i \leq n$ for the component of clause $\mathcal{C}^j$ in any red blue graph $G(\pi, \mathcal{S})$. 
\end{lemma}

\begin{lemma}\label{unsatdisconn}
For any red blue graph $G(\pi, \mathcal{S})$, the blue component containing $q$ symbols of any clause $\mathcal{C}^j$ can be connected to the graph $G(\pi, \mathcal{S})$ only via a red edge of type $(p_i^j, q_i^j)$, (or $(\bar{p}_i^j, \bar{q}_i^j)$) for $1 \leq j \leq m$, and $1\leq i \leq n$, without disconnecting another blue component from $G(\pi, \mathcal{S})$. 
\end{lemma}

%\newpage
\begin{lemma}\label{left}
Max-$3SAT(\Phi) \leq m-c$ $\Longrightarrow$ $bs(\pi) \geq 6m+2n-1+c$.
\end{lemma}

\proof

\noindent
Given a satisfying assignment of $\Phi$, if $\mathcal{C}^j$ is unsatisfied in $\Phi$, by Lemma~\ref{unsat}, the blue component with $q$ symbols of $\mathcal{C}^j$ will be disconnected in any $G(\pi, \mathcal{S})$. Therefore when Max-$3SAT(\Phi) \leq m-c$, for $c$ such unsatisfied clauses, we would have $c$ such disconnected blue components with $q$ symbols in any red blue graph $G(\pi, \mathcal{S})$. If we try to connect any of these disconnected blue components with $q$ symbols via any other red edge than type $(p_i^j, q_i^j)$, (or $(\bar{p}_i^j, \bar{q}_i^j)$) for $1\leq i \leq n$, it will disconnect at least $1$ other blue component from $G(\pi, \mathcal{S})$ by Lemma~\ref{unsatdisconn}. This proves when Max-$3SAT(\Phi) \leq m-c$, we have at least $c$ disconnected blue components {\em for any red blue graph} $G(\pi, \mathcal{S})$. And therefore by Theorem~\ref{tighterlb}, we will have $bs(\pi) \geq 6m+2n-1+c$. 
\qed
%\noindent
%For the complete proof of Lemma~\ref{left}, please see Appendix~\ref{sec3}.

\begin{lemma}\label{right}~\cite{blocksort03}
Max-$3SAT(\Phi)=m$ $\Longrightarrow$ $bs(\pi)=6m+2n-1$.
\end{lemma}
\proof This was already proved in~\cite{blocksort03}. The outline is at least one literal in each clause in $\Phi$ is true. Hence we can find all the $m$ red edges of type $(p_i^j, q_i^j)$ mentioned Corollary~\ref{tree1}. 
\qed

\begin{lemma}\label{apx}
It is $\NP$-Hard to approximate \textsc{Block Sorting} to within a factor of $1.02$.
\end{lemma} 

\proof
Taking $c=\frac{m}{8}$, we have from Lemma~\ref{left}, Max-$3SAT(\Phi) \leq \frac{7m}{8}$ $\Longrightarrow$ $bs(\pi) \geq 6m+2n-1+\frac{m}{8}$. This proves the lemma.
\qed

\noindent
Lemma~\ref{apx} leads to Theorem~\ref{hard}.

\begin{theorem}\label{hard}
\textsc{Block Sorting} is Max-$\mathcal{SNP}$-Hard (APX-Hard).
\end{theorem}

\noindent
We now state and prove the following lemmas, which lead to Lemma~\ref{unsat}, and Lemma~\ref{unsatdisconn}.
%\newpage

\begin{lemma}\label{joinpairs}
In $\pi$, the pairs $(u_i,\upsilon_i)$ $\forall 1 \leq i \leq n$, and $(\ell^j,r^j)$ $\forall 1 \leq j \leq m+n$ can always be joined to form blocks before they are moved in any block sorting schedule $\mathcal{S}$ on $\pi$. 
\end{lemma}
\noindent
We define a set of red edges $\mathbf{E}=\{ (u_i, \upsilon_i)$ $\forall 1 \leq i \leq n,$ $(\ell^j, r^j)$ $\forall 1 \leq j \leq m+n,$ and $(s, \ell^1)\}$ for any red blue graph $G(\pi, \mathcal{S})$.
\begin{lemma}\label{num-red-edge}
In any red-blue graph $G(\pi, \mathcal{S})$, no red edges in the set $\mathbf{E}$ cross each other . Hence the number of red edges which could be drawn in any $G(\pi, \mathcal{S})$ is at least $m+2n+1$.
\end{lemma}

\begin{corollary}\label{easyproof}
There exists a block sorting schedule $\mathcal{S}$ for $\pi$ of length $7m+2n-1$ steps.
\end{corollary}

\noindent
The $|\mathbf{E}|=m+2n+1$ red edges connect the respective blue components to each other. Therefore $m+2n+2$ blue components of $G(\pi, \mathcal{S})$ get connected to each other by the set of edges $\mathbf{E}$. The blue components that do not get connected by the edges of $\mathbf{E}$ are  the blue components with the $q$ symbols. Lemma~\ref{vimp} is a direct implication of Claim A, B, and C of the proof of Lemma $11$ of~\cite{blocksort03}.

\begin{lemma}\cite{blocksort03}\label{vimp}
In any red blue graph $G(\pi, \mathcal{S})$ the set $\mathbf{E}$ of is the only set of non crossing red edges, which can connect all the $m+2n+2$ blue components of $G(\pi, \mathcal{S})$ to which they belong to. So any edge absent from $\mathbf{E}$ in $G(\pi, \mathcal{S})$ would imply at least one disconnected blue component in $G(\pi, \mathcal{S})$. 
\end{lemma}

\begin{lemma}\label{joinpairspq}
At most one pair $(p_i^j, q_i^j)$ (or $(\bar{p}_i^j, \bar{q}_i^j)$) from each clause encoding $1\leq j \leq m$, and each variable $1\leq i \leq n$ can be joined to form blocks before they are moved. Moreover, if $(p_i^j, q_i^j)$ are joined before they are moved for any clause $1 \leq j \leq m$, then $(\bar{p}_i^k, \bar{q}_i^k)$ cannot be joined before they are moved for any clause $1 \leq k \leq m$, and $1\leq i \leq n$, and $k \neq j$ in any block sorting schedule $\mathcal{S}$ on $\pi$.  
\end{lemma}

%\noindent
%Given any block sorting schedule $\mathcal{S}$ on $\pi$, we construct the red-blue graph $G(\pi, \mathcal{S})$. 

%This explains how we can save at most $m$ moves and $bs(\pi) \geq 6m+2n-1$.
\begin{corollary}\label{tree1}
In any red-blue graph $G(\pi, \mathcal{S})$, there can be at most $m$ red edges between the pairs $(p_i^j, q_i^j)$ (or $(\bar{p}_i^j, \bar{q}_i^j)$) for $1\leq i \leq n$, and $1\leq j \leq m$, one such red edge in each clause encoding. Furthermore, if there is a red edge between $(p_i^j, q_i^j)$ in a clause encoding for $1\leq i \leq n$ and, $1\leq j \leq m$, there cannot be a red edge between $(\bar{p}_i^k, \bar{q}_i^k)$ in any clause encoding for $1\leq i \leq n$ and, $1\leq k \leq m$, $k \neq j$. 
\end{corollary}

\section{A New Lower Bound for {\sc Block Sorting}}\label{sec5}
\noindent
The problem {\sc Block Merging} has been introduced in~\cite{blocksort06}. It is defined as follows:
\begin{center}
\framebox{
\begin{tabular}{l}
\noindent {\sc Block Merging Problem}
\\ \noindent {\sc Input:} A multiset $\mathbb{S} = \{S_1, S_2, \cdots , S_l\}$ of disjoint increasing sequences 
\\whose union is $[n]$, an integer $m$.
\\ \noindent {\sc Output:} The multiset $\mathbb{M}_n = \{id_n, \epsilon, \cdots , \epsilon\}$.
\\ \noindent {\sc Constraint:} A block is allowed to be moved if it is contained in
\\{\em at most one increasing sequence} $S_i$.
\\ \noindent {\sc Question:}  Is $bm(\mathbb{S}) \leq m$? 
\end{tabular}
}
\end{center}
\noindent
The block merging distance $bm(\mathbb{S})$ is the minimum number of block moves to transform $\mathbb{S}$ to $\mathbb{M}_n$. A block move is defined as: Pick a block from any sequence $S_i$, and insert it into some other sequence $S_j$ so that it merges with a block there. A block is allowed to be moved if it is contained in {\em at most one increasing sequence} $S_i$. Any permutation $\pi$ can easily be decomposed into the multiset of maximal increasing subsequences $\mathbb{S_{\pi}}$, which could be an input to {\sc Block Merging }. Lemma~\ref{pbm} and~\ref{lbm} have been proved in~\cite{blocksort06}.

\begin{lemma}\label{pbm}
{\sc Block Merging} $\in$ {\bf P}.
\end{lemma}

\begin{lemma}\label{lbm}
For any $\pi$, $bm(\mathbb{S_{\pi}}) \geq bs(\pi) \geq \frac{bm(\mathbb{S_{\pi}})}{2}$.
\end{lemma}

\noindent
Lemma~\ref{lbm} gives a new lower bound for {\sc Block Sorting}. It also says that {\sc Block Merging} approximates {\sc Block Sorting} by a Factor of $2$. Lemma~\ref{pbm} along with Lemma~\ref{lbm} gives polynomial factor $2$ approximation algorithm for {\sc Block Sorting}. All the omitted proofs of this section appear in Appendix~\ref{sec4}. 

\noindent
We relax the constraint for {\sc Block Merging} and define the problem {\sc $k$-Block Merging}. Specifically in {\sc $k$-Block Merging}, we are allowed to move a block which is contained in {\em at most $k$ increasing sequence}. We are done when we have at most $k$ sequences whose concatenation gives the identity permutation $id_n$. Formally, we define {\sc $k$-Block Merging} as:

\begin{center}
\framebox{
\begin{tabular}{l}
\noindent {\sc $k$-Block Merging Problem}
\\ \noindent {\sc Input:} A multiset $\mathbb{S} = \{S_1, S_2, \cdots , S_l\}$ of disjoint increasing sequences
\\ whose union is $[n]$, integers $m$, and $k \leq l$.
\\ \noindent {\sc Output:} The multiset $\mathbb{M}_n = \{S^{\prime}_1, S^{\prime}_2, \cdots , S^{\prime}_k, \epsilon, \cdots , \epsilon\}$ 
\\such that $S^{\prime}_1S^{\prime}_2 \cdots S^{\prime}_k=id_n$.
\\ \noindent {\sc Constraint:} A block is allowed to be moved if it is contained in
\\{\em at most $k$ increasing sequence} $S_i$.
\\ \noindent {\sc Question:}  Is $k$-$bm(\mathbb{S}) \leq m$? 
\end{tabular}
}
\end{center}

\noindent
$k$-$bm(\mathbb{S})$ is the number of $k$-block merging moves to transform $\mathbb{S}$ to $\mathbb{M}_n$.

\begin{lemma}\label{nub}
$bs(\pi) \leq$ $k$-$bm(\mathbb{S}_{\pi}) \leq bm(\mathbb{S}_{\pi})$.
\end{lemma}

\begin{lemma}\label{nlb}
$bs(\pi) \geq  \frac{k\mbox{-}bm(\mathbb{S}_{\pi})}{1+ \frac{1}{k}}$.
\end{lemma}
\proof
For any permutation $\pi$, given a block sorting schedule $b_1, b_2, \cdots, b_m$ of $m$ moves we need to prove that we can have a $k$-block merging schedule of at most $m(1+\frac{1}{k})$ moves. For any block sorting move $b_i$, if $b_i$ moves the block $B \in \pi$, we move $B$ in the corresponding $k$-block merging move if it is contained in at most $k$ increasing sequences in $S_{\pi}$. Else, if $B$ is contained in $x>k$ increasing sequences, we perform $\left\lfloor\frac{x-1}{k} \right\rfloor$ moves to get $B$ in at most $k$ increasing sequences. Next we move $B$. This way, each move in the block sorting sequence, can be performed by one or more moves in the $k$-block merging sequence. 

\noindent 
We define two sets $Inc(\pi)=\{ \pi_i | \pi_i < \pi_{i+1}, \forall 1 \leq i < n\}$, and $Inc(\mathbb{S})=\{  i \in [n-1]|i$ is not the last element of any subsequence $\mathbf{S} \in \mathbb{S}\}$. In other words, $Inc(\pi)$ is the set of elements of $\pi$ which are lesser than their immediate successor element in $\pi$. We have $Inc(id_n)=Inc(\mathbb{M}_n)=[n-1]$. At the beginning, for any $\pi$, we have $Inc(\pi)=Inc(\mathbb{S}_{\pi})$. But as we execute a block sorting, and its corresponding $k$-block merging schedule, $Inc(\mathbb{S}_{\pi}^{i}) \subseteq Inc(\pi^{i})$ at any step $i$. The reason is, a block $B$ can be fragmented at step $i$, and that would make $Inc(\mathbb{S}_{\pi})$ not contain elements which are in $Inc(\pi)$. Hence, the defragmentation steps that we perform for $k$-block merging, actually decreases the difference between $Inc(\mathbb{S}_{\pi}^{i})$ and $Inc(\pi^{i})$ for step $i$.

\noindent
Let $c_i$ be the actual cost of the corresponding $k$-block merging moves performed for a single block sorting move at step $i$. Then $c_i = 1 + \left\lfloor\frac{x-1}{k} \right\rfloor$. We perform an amortized cost analysis of the amortized cost $a_i$ for each step, such that $a_i \geq c_i$, $\forall$ $i$. Then we bound $a_i$ by $1+\frac{1}{k}$ $\forall$ $i$. For step $i$, let $\mathbf{P}_i=Inc(\pi^{i})$ and $\mathbf{Q}_i=Inc(\mathbb{S}_{\pi}^{i})$. Further, let $|\mathbf{P}_i|=p_i$, and $|\mathbf{Q}_i|=q_i$. We define potential function $\phi_i=\frac{p_i-q_i}{k}$. Then the amortized cost for step $i$ becomes $a_i = c_i+ \phi_i - \phi_{i-1}$.

\noindent
To complete the proof, we need to show that $\phi_i - \phi_{i-1} \leq \frac{1}{k}-\frac{x-1}{k}$. We have $\phi_i - \phi_{i-1} = \frac{(p_i-q_i)-(p_{i-1}-q_{i-1})}{k} = \frac{(p_i-p_{i-1})-(q_i-q_{i-1})}{k}$. We calculate the change in $p_i-p_{i-1}$, and $q_i-q_{i-1}$ for all the $k$-block merging moves of step $i$. Let the block $B$ be moved to its predecessor block $A$ by the block move $b_i$ (the other case is analogous).

\noindent
We first find the bound on the value of $p_i-p_{i-1}$. Let the first and last elements of the block $B$ be $b$ and $c$ respectively, and the last element of block $A$ be $e$, that is $e=b-1$. It is clear that $e\in\mathbf{P}_i$, since $e<b$. Further, $c\in\mathbf{P}_i \Leftrightarrow e\in\mathbf{P}_{i-1}$. Now we observe that $p_i-p_{i-1}=1$ if 
\begin{enumerate}
\item either $a\notin\mathbf{P}_{i-1}$ {\em and} $e\notin\mathbf{P}_{i-1}$, 
\item or if {\em exactly one among} $a$ and $c$ $\in\mathbf{P}_{i-1}$, {\em and} $a\in\mathbf{P}_{i}$.
\end{enumerate}
\noindent
In every other case $p_i-p_{i-1}=0$.

\noindent
Consider the $k$-block merging moves on to simulate block move $b_i$. Recall that the block $B$ is fragmented within $x$ increasing sequences. We need to make at most $\frac{x-1}{k}$ moves to bring $B$ into at most $k$ increasing subsequences. Then we move $B$ to its predecessor $A$, as done by block move $b_i$. Since $k\geq1$, the maximum number of such defragmentation moves performed here is $x-1$. These $x-1$ moves can contribute at most $x-1$ to $q_i-q_{i-1}$. In fact they contribute $x-1$ if $c\notin\mathbf{Q}_{i-1}$, else they contribute $x-2$. Again the block move to move $B$ to $A$ adds either of $e$ or $c$ to $\mathbf{Q}_i$ and hence contributes $1$ to $q_i-q_{i-1}$. But we have $a\notin\mathbf{Q}_{i}$. Hence the overall contribution by the block move is $1$ if $a\notin\mathbf{Q}_{i-1}$, else its $0$. 

\noindent
To sum it up, block move $b_i$ increases  $p_i-p_{i-1}$ by $0$ or $1$, and the equivalent $k$-block merging moves increase $q_i-q_{i-1}$ by either at most $x-1$ or at most $x-2$. In the latter case, we have both $a$ and $c\in \mathbf{Q}_{i-1}$, and hence $a$ and $c\in \mathbf{P}_{i-1}$ which means $p_i-p_{i-1}=0$. 

\noindent
Therefore we have $\phi_i - \phi_{i-1} = \frac{(p_i-p_{i-1})-(q_i-q_{i-1})}{k} \leq \frac{1-(x-1)}{k} = \frac{1}{k} - \frac{x-1}{k}$. 
\qed

\noindent
Lemma~\ref{nub}, and~\ref{nlb} lead to Theorem~\ref{napx}. Theorem~\ref{hard} and ~\ref{napx} lead to Corollary~\ref{hard2}.
\begin{theorem}\label{napx}
{\sc $k$-Block Merging} approximates {\sc Block Sorting} by a factor of $1+\frac{1}{k}$.
\end{theorem}

\begin{corollary}\label{hard2}
{\sc $k$-Block Merging} is $\NP$-Hard for $k>48$.
\end{corollary}

\noindent
Lemma~\ref{nlb} gives us a new lower bound for {\sc Block Sorting} via {\sc $k$-Block Merging}. In~\cite{blocksort06}, given any $\pi$, and $\mathbb{S}_{\pi}$, a directed graph $G=(V,E)$ has been constructed such that $V=[n]$, and $(u,\upsilon)\in E$ if $u$ and $\upsilon$ belong to the same increasing subsequence in $\pi$. Two edges $(i,j)$, and $(k,l)$ {\em cross} each other if $i \leq k<j \leq l$, or $k \leq i<l \leq j$. A set $E^{\prime} \subseteq E$ is called a {\em non-crossing set} if no two edges of $E^{\prime}$ cross. The size of a largest non-crossing in $\mathbb{S}_{\pi}$ is denoted by $c(\mathbb{S}_{\pi})$. Lemma~\ref{bm} has been proved in~\cite{blocksort06}.

\begin{lemma}\label{bm}
$bm(\mathbb{S}_{\pi})=n-1-c(\mathbb{S}_{\pi})$.
\end{lemma}

\begin{lemma}\label{nbm}
$k$-$bm(\mathbb{S}_{\pi}) \geq \frac{n-1-c(\mathbb{S}_{\pi})}{k}$.
\end{lemma}

\begin{corollary}\label{coro}
$k$-$bm(\mathbb{S}_{\pi}) \geq \frac{bm(\mathbb{S}_{\pi})}{k}$.
\end{corollary}

\noindent
Corollary~\ref{coro} tells us that the polynomial time algorithm of~\cite{blocksort06} for {\sc Block Merging} is actually a $k$-approximation algorithm for {\sc $k$-Block Merging}. We know {\sc $k$-Block Merging} to be polynomial time solvable for $k=1$ from~\cite{blocksort06}, and our results prove it to be $\NP$-Hard for $k>48$. But we do not know whether it is polynomial time solvable for $k=2$. If it is, then we would have a $1.5$-approximation algorithm for {\sc Block Sorting}. But if it is not, then {\sc Block Sorting} would be inapproximable to within a factor of $1.5$. It is still open whether we can design an algorithm  with an approximation ratio better than $2$ for {\sc Block Sorting}.
\subsubsection*{Acknowledgments.} The second author thanks his mentor Atri Rudra for being a constant source of encouragement and inspiration. 

%\section{Conclusion}

%We have shown the problem {\sc Sorting by $k$-Block Interchanges} to be hard for $k=1$. It would be interesting to know what happens if we relax the constraint a bit. What if one of the sub-strings exchanged be a block as defined in this work while the other be any sub-string of the given permutation? Another interesting problem would be to find whether the problem is Max-$\mathcal{SNP}$-complete, which we conjecture would depend on whether or not {\sc Block Sorting} is Max-$\mathcal{SNP}$-complete.  

%\newpage

\pagebreak

\appendix

\section{A Few Definitions}\label{glossary}

\subsection{$\APX$}
A problem is said to belong to class $\APX$ if we can design a constant factor approximation algorithm for it. By definition $\APX \subset \NP$.   

\subsection{Polynomial Time Approximation Scheme (PTAS)}
A PTAS is an algorithm which takes an instance of an optimization problem and a parameter $\varepsilon > 0$ and, in polynomial time, produces a solution that is within a factor $1 + \varepsilon$ of being optimal (or $1 - \varepsilon$ for maximization problems). For example, for the Euclidean traveling salesman problem, a PTAS would produce a tour with length at most $(1 + \varepsilon)L$, with $L$ being the length of the shortest tour.

\subsection{$\APX$-hardness}
A problem is said to be $\APX$-Hard, if we cannot design any PTAS for it unless $\Poly = \NP$.

\section{An Example of a Block Move and a Block Sorting Schedule}\label{sortingex}

\begin{figure}\begin{center}
\includegraphics[scale=1]{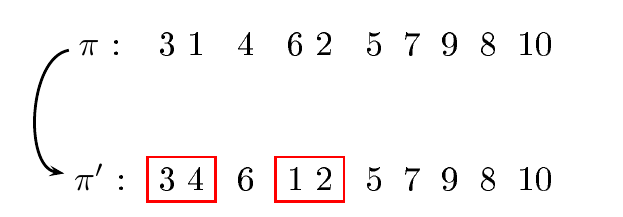}
\caption{\label{ex1} An example of a block move}
\end{center}
\end{figure}

\begin{figure}\begin{center}
\includegraphics[scale=.7]{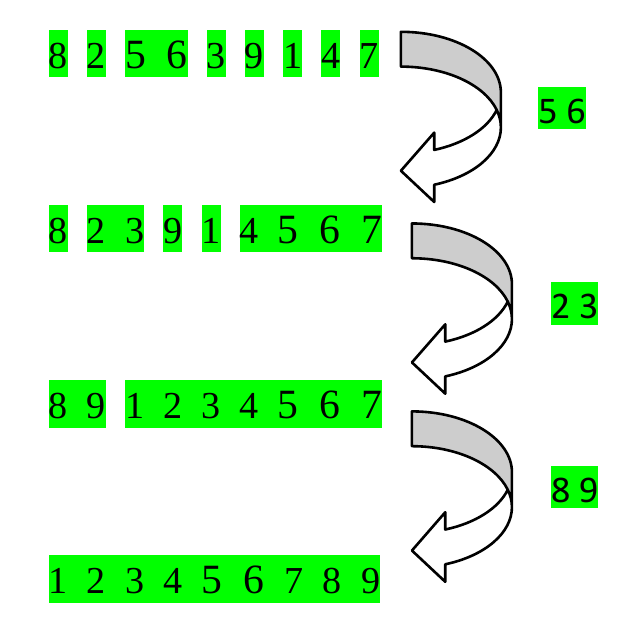}
\caption{\label{example:schedule} An example of a block sorting schedule}
\end{center}
\end{figure}

\noindent
A block sorting schedule is shown on permutation 8 2 5 6 3 9 1 4 7 in Figure~\ref{example:schedule}. The block moves are indicated at each step. 

\pagebreak

\section{An Example of the Reduction Procedure}\label{exred}
\begin{figure}\begin{center}
\includegraphics[scale=.96]{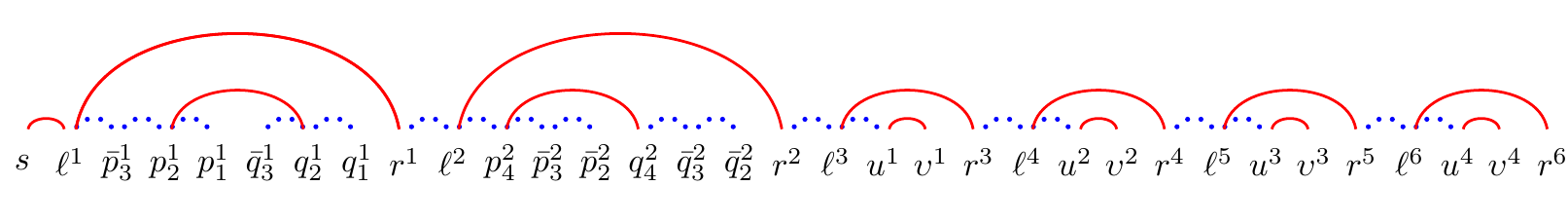}
\caption{\label{reduction} An example of the reduction for $\Phi = (\bar{x}_3 \vee x_2 \vee x_1)\wedge(x_4 \vee \bar{x}_3 \vee \bar{x}_2)$}
\end{center}
\end{figure}

A red-blue graph for the $\pi$ in Figure~\ref{reduction} has been drawn taking $x_2=1$, and $x_4=1$. Since $\Phi$ gets satisfied by this assignment, this red-blue graph is a tree. This signifies that the corresponding block sorting schedule is perfect.

\pagebreak
\section{An Example of an Unsatisfied Clause in $\Phi$}\label{exunsat}
\begin{figure}\begin{center}
\includegraphics[scale=.75]{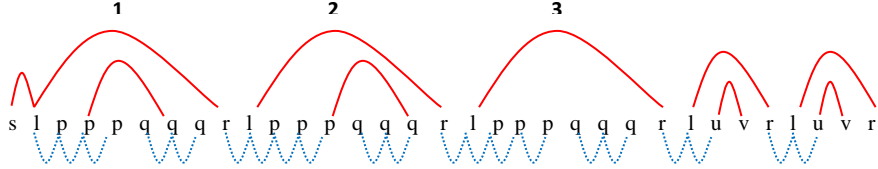}
\caption{\label{example} An example of the encoding of an unsatisfied clause.}
\end{center}
\end{figure}

\section{Omitted Proofs from Section~\ref{redblue}}\label{sec2}

\subsection{Proof of Lemma~\ref{disconn}}
From~\cite{blocksort03}, we know the following:
\begin{enumerate}
\item For any block sorting schedule $\mathcal{S}$ on any $\pi$, if $m$ is the number of moves in $\mathcal{S}$, $\#blocks(\pi)$ is the number of blocks in $\pi$, and $\#red(G(\pi, \mathcal{S}))$ is the number of red edges of $G(\pi, \mathcal{S})$, then $m=\#blocks(\pi)-1-\#red(G(\pi, \mathcal{S}))$.
\item $G(\pi, \mathcal{S})$ for any $\pi$ and any $\mathcal{S}$ on $\pi$ is acyclic.
\end{enumerate}
The second property implies, that if there are $b$ blue components in any $G(\pi, \mathcal{S})$, then the number of red edges $\#red(G(\pi, \mathcal{S}))$ is at most $b-1$. Therefore we have $m \geq \#blocks(\pi)-b$. If we have $d$ disconnected blue components in $G(\pi, \mathcal{S})$, then we have at most $b-1-d$ red edges in $G(\pi, \mathcal{S})$. By the first property in this case we have $m \geq \#blocks(\pi)-b+d$, and $\#blocks(\pi)-b=rev(\pi)$. This proves the lemma.
\qed 

%\appendix
\section{Omitted Proofs from Section~\ref{npc}}\label{sec3}

\subsection{Proof of Corollary~\ref{easyproof}}
This follows from the fact that we can always have the edges in $\mathbf{E}$ in any $G(\pi, \mathcal{S})$. Therefore, we can always construct a block sorting schedule of length $7m+2n-1$ in the following way:
\begin{enumerate}
\item Move all the term symbols to their proper places in any order. This would take $6m$ moves (one move for each symbol).
\item Now move the pairs $u_i\upsilon_i$ to their proper places. Note that they have already formed blocks by the above step. This takes $n$ steps.
\item All the $\ell^jr^j$ pairs have formed blocks, but are in the reverse order after the above two steps. So getting them in order requires $m+n-1$ block moves.\qed
\end{enumerate}

\subsection{Proof of Lemma~\ref{one-on-one}}
This is already implied from the construction of permutation $\pi$, from formula $\Phi$. Specifically, given a satisfying assignment for $\Phi$, it satisfies $k$ clauses say $\mathcal{C}^1$ to $\mathcal{C}^k$, {\em if and only if} we have a red blue graph $G(\pi, \mathcal{S})$ for which we have a set of $k$ non-crossing edges $\{ (p_i^j, q_i^j)$ or $(\bar{p}_i^j, \bar{q}_i^j), 1\leq i \leq n, 1 \leq j \leq k $ {\em such that we have exactly $1$ edge from the component of each clause} $\mathcal{C}^j$ {\em in} $G(\pi, \mathcal{S})$ $\}$. To prove what we have just stated, we observe when clauses $\mathcal{C}^1$ to $\mathcal{C}^k$ are satisfied, we have at least one true literal for each clause. We pick one true literal $x_i^j$ (or $\bar{x}_i^j$) for each clause $\mathcal{C}^j$, $1 \leq j \leq k$. The $k$ red edges $(p_i^j, q_i^j)$ or $(\bar{p}_i^j, \bar{q}_i^j)$ corresponding to $x_i^j$ (or $\bar{x}_i^j$) for each clause $\mathcal{C}^j$ do not cross. For the other direction, if we have $k$ non-crossing red edges in any $G(\pi, \mathcal{S})$, one from each clause component, then we can set the corresponding literal to be true for that clause and satisfy those $k$ clauses. This property of $\pi$ was used along with other properties in~\cite{blocksort03}, to prove the $\NP$-hardness of {\sc Block Sorting}.\qed

\subsection{Proof of Lemma~\ref{unsat}}
Since $\mathcal{C}^j$ is unsatisfied in $\Phi$ for all variables $z_i = (x_i or, \bar{x_i}) \in \mathcal{C}^j$, $\forall 1\leq i \leq n$, we have $\bar{z}_i$, the complement of $z_i$ true in at least another clause $\mathcal{C}^k$, and $\bar{z}_i$ satisfies clause $\mathcal{C}^k$. Therefore by Lemma~\ref{joinpairspq}, any pair $(p_i^j, q_i^j)$ $\forall 1 \leq i \leq n$, in the component of clause $\mathcal{C}^j$ will cross with at least one pair $(\bar{p}_i^k, \bar{q}_i^k)$ $\forall 1 \leq i \leq n$, in the component of clause $\mathcal{C}^k$, for $1\leq k \leq m$, and $k \neq j$. By Corollary~\ref{tree1}, any red edge $(p_i^j, q_i^j)$  $\forall 1 \leq i \leq n$ would cross at least cross another red edge of the form $(\bar{p}_i^k, \bar{q}_i^k)$ for $1\leq k \leq m$, and $k \neq j$ in any red blue graph $G(\pi, \mathcal{S})$. Since we have all the other red edges (for the components other than that of clause $\mathcal{C}^j$) in $G(\pi, \mathcal{S})$, we cannot have any red edge drawn between any pair $(p_i^j, q_i^j)$ for the clause component of $\mathcal{C}^j$.\qed

\subsection{Proof of Lemma~\ref{unsatdisconn}}
\noindent
We prove that for any red blue graph $G(\pi, \mathcal{S})$, the blue component containing $q$ symbols of any clause $\mathcal{C}^j$ can be connected to the graph $G(\pi, \mathcal{S})$ only via a red edge of type $(p_i^j, q_i^j)$, (or $(\bar{p}_i^j, \bar{q}_i^j)$) for $1 \leq j \leq m$, and $1\leq i \leq n$. Any other red edge drawn to any of the symbols $q_i^j$ of the blue component, will cross a red edge $\in \mathbf{E}$ of Lemma~\ref{num-red-edge} leading to its removal from $G(\pi, \mathcal{S})$. This disconnects another blue component from $G(\pi, \mathcal{S})$ by Lemma~\ref{vimp}. We further show that any red edges other than type $(p_i^j, q_i^j)$, (or $(\bar{p}_i^j, \bar{q}_i^j)$) drawn to the blue components with $q$ symbols for different clauses $\mathcal{C}^j$, $1 \leq j \leq m$ would cross different red edges $\in \mathbf{E}$. 

\noindent
We now exhaustively consider all other red edges to connect the blue component with $q$ symbols to $G(\pi, \mathcal{S})$. We show that each of these red edges are either non-existent, or cross with another red edge $\in \mathbf{E}$, hence disconnect another blue component from $G(\pi, \mathcal{S})$. We will prove things for $(p_i^j, q_i^j)$ and without loss of generality will omit $(\bar{p}_i^j, \bar{q}_i^j)$.

\begin{enumerate}

\item {\bf There cannot be any red edge from $s$ to $q_i^j$, any $r^k$ to $q_i^j$, $\ell^j$ to $q_i^j$, and from $q_i^j$ to $u_i$ for $1\leq i \leq n$, $1 \leq k < j \leq m$.} This is true because $s>q_i^{j}$, $r^k>q_i^{j}$, $\ell^j>q_i^{j}$, and $q_i^j>u_i$ $\forall 1\leq i \leq n, 1\leq k < j \leq m$. And $\pi(s)<\pi(q_i^{j})$, $\pi(r^k)<\pi(q_i^{j})$, $\pi(\ell^j)<\pi(q_i^{j})$, and $\pi(q_i^j)<\pi(u_i)$ $\forall 1\leq i \leq n, 1\leq k < j \leq m$. Hence the pairs $(s, q_i^j)$, $(r^k, q_i^j)$, $(\ell^j, q_i^j)$, and $(q_i^j, u_i)$ $\forall 1\leq i \leq n, 1\leq k < j \leq m$, are not in order.

\item {\bf A red edge cannot be drawn between pairs $(p_i^j, q_k^j)$ from each clause encoding $1\leq j \leq m$, and variables $1\leq i \leq n$, $1\leq k \leq n$ for $i \neq k$, without disconnecting a blue component.} The pairs $(p_i^j, q_k^j)$ for $1 \leq k<i \leq n$ and $1\leq j \leq m$ are not in order, since $p_i^j > q_k^j$, and $\pi(p_i^j) < \pi(q_k^j)$  for $1 \leq k<i \leq n$ and $1\leq j \leq m$. 

\noindent
For $1 \leq i<k \leq n$ and $1\leq j \leq m$, $u_i<p_i^j<\upsilon_i<q_k^j$ and $p_i^j<u_k<q_k^j<\upsilon_k$. Hence the red edges $(p_i^j, q_k^j)$ for $1 \leq i<k \leq n$ and $1\leq j \leq m$ cross with both the red edges $(u_i, \upsilon_i)\in \mathbf{E}$, and $(u_k, \upsilon_k) \in \mathbf{E}$ by Lemma~\ref{red3}. 

\item {\bf A red edge cannot be drawn from any $q_i^j$ to any $r^k$ for $1\leq i \leq n$, $1 \leq j \leq k \leq m$, without disconnecting a blue component.} 
We have $u_i<q_i^j<\upsilon_i<r^k$ for  $1\leq i \leq n$, $1 \leq j \leq k \leq m$. Hence red edges $(q_i^j, r^k)$ and $(u_i, \upsilon_i) \in \mathbf{E}$ cross by Lemma~\ref{red3} for  $1\leq i \leq n$, $1 \leq j \leq k \leq m$. Therefore drawing red edge $(q_i^j, r^k)$ disconnects the blue component with the $\upsilon_i$ symbol from the graph. 

\noindent
Moreover, the red edge $(q_i^j, r^k)$ for any $1\leq i \leq n$, $1 \leq j \leq k \leq m$, also cross with red edge $(\ell^j, r^j) \in \mathbf{E}$ for that $j$, by Lemma~\ref{red1}. Hence drawing a red edge of type $(q_i^j, r^k)$ for $1\leq i \leq n$, $1 \leq j \leq k \leq m$, disconnects more than one blue component from the graph.

\item {\bf A red edge cannot be drawn from any $q_i^j$ to any $\ell^k$ for $1\leq i \leq n$, $1 \leq j < k \leq m$ without disconnecting a blue component.}
We have $u_i<q_i^j<\upsilon_i<\ell^k$ for  $1\leq i \leq n$, $1 \leq j < k \leq m$. Hence red edges $(q_i^j, \ell^k)$ and $(u_i, \upsilon_i) \in \mathbf{E}$ cross by Lemma~\ref{red3} for  $1\leq i \leq n$, $1 \leq j < k \leq m$. Therefore drawing red edge $(q_i^j, \ell^k)$ disconnects the blue component with the $\upsilon_i$ symbol from the graph. 

\noindent
Again we have $\pi(\ell^j)<\pi(q_i^j)<\pi(r^j)<\pi(\ell^k)$ for  $1\leq i \leq n$, $1 \leq j < k \leq m$. Hence red edges $(q_i^j, \ell^k)$ and $(\ell^j, r^j) \in \mathbf{E}$ for  $1\leq i \leq n$, $1 \leq j < k \leq m$, cross by Lemma~\ref{red2}. Hence drawing a red edge of type $(q_i^j, \ell^k)$ for $1\leq i \leq n$, $1 \leq j < k \leq m$, disconnects more than one blue component from the graph.

\item {\bf A red edge cannot be drawn from any $q_i^j$ to any $\upsilon_k$ for $1\leq i \leq n$, $1 \leq j \leq m$, and $1 \leq k \leq m$ without disconnecting a blue component.} For $k<j$, we have $q_i^j>\upsilon_k$, and $\pi(q_i^j)<\pi(\upsilon_k)$. Hence the pair $(q_i^j,\upsilon_k)$ is out of order for $k<j$.

\noindent
For $k \geq j$, if we draw a red edge from $q_i^j$ to $\upsilon_k$, the red edges $(q_i^j,\upsilon_k)$ and $(u_k,\upsilon_k) \in \mathbf{E}$ cannot coexist by Lemma~\ref{red1}. Again we have $\pi(q_i^j)<\pi(\ell^{m+k})<\pi(\upsilon_k)<\pi(r^{m+k})$. Hence red edges $(q_i^j,\upsilon_k)$, $(\ell^{m+k},r^{m+k}) \in \mathbf{E}$ cross by Lemma~\ref{red2}. Hence again more than one blue components get disconnected in this case. The component with the $\upsilon_k$ symbol, and the component with the $r^{m+k}$ symbol.
\end{enumerate}
Red edges other than type $(p_i^j, q_i^j)$ drawn to connect the blue component with symbol $q_i^j$ for different clauses $\mathcal{C}^j$ cross different edges of the set $\mathbf{E}$, as stated earlier.
\qed

\subsection{Proof of Lemma~\ref{joinpairs}}

We need to show the following for all the pairs mentioned in the lemma:

\begin{enumerate}
\item They are in order: $a<b$, and $\pi(a)<\pi(b)$, $\forall (a,b)$.
\item Any two pairs do not violate any conditions of Lemmas~\ref{red1}, ~\ref{red2}, ~\ref{red3}, or ~\ref{red4}. That is, they do not cross each other. 
\end{enumerate}

\noindent
To recall, the boolean formula $\Phi$ has $n$ variables, and $m$ clauses. We have $\ell^j<r^j$, and $\pi(\ell^j)<\pi(r^j)$ $\forall 1 \leq j \leq m+n$. Hence the pairs $(\ell^j, r^j)$ are in order $\forall 1 \leq j \leq m+n$. Further $u^k<\upsilon^k$, and $\pi(u^k)<\pi(\upsilon^k)$ $\forall 1 \leq k \leq n$. Hence the pairs $(u^k,\upsilon^k)$ are in order $\forall 1 \leq k \leq n$. We state and prove the following claim:

\begin{claim}\label{crossing}
For any $1 \leq i \leq m+n, 1 \leq j \leq m+n , 1 \leq k \leq n$ the pair $(\ell^i, r^i)$ does not cross the pairs $(\ell^j, r^j)$, and $(u^k,\upsilon^k)$  
\end{claim}

\proof

\begin{enumerate}
\item $\ell^i<r^i<\ell^j<r^j$ $\forall 1 \leq j<i \leq m+n$ and, $\pi(\ell^i)<\pi(r^i)<\pi(\ell^j)<\pi(r^j)$ $\forall 1 \leq i<j \leq m+n$. Hence the pairs $(\ell^i, r^i)$, and $(\ell^j, r^j)$ do not cross $\forall 1 \leq i \leq m+n , 1 \leq j \leq m+n $.
\item $u^k<\upsilon^k<\ell^j<r^j, \forall 1 \leq k \leq n,1 \leq j \leq m+n$, and $\pi(\ell^j)<\pi(u^k)<\pi(\upsilon^k)<\pi(r^j), \forall 1 \leq k \leq n, j=m+k$.
\item  $\pi(\ell^j)<\pi(r^j)<\pi(u^k)<\pi(\upsilon^k),  \forall 1 \leq j \leq n, j<m+k$, and $\pi(u^k)<\pi(\upsilon^k)<\pi(\ell^j)<\pi(r^j), \forall 1 \leq k \leq n, j>m+k$.   
\end{enumerate}
\noindent
$2.$ and $3.$ imply, the pairs $(u^k,\upsilon^k)$, and $(\ell^j, r^j)$ do not cross each other $\forall 1 \leq k \leq n , 1 \leq j \leq m+n $.

Claim~\ref{crossing} completes the proof of Lemma~\ref{joinpairs}. \qed

\subsection{Proof of Lemma~\ref{num-red-edge}}
With a similar argument as in the proof of Lemma~\ref{joinpairs}, we can show that the pair $(s, \ell^1)$ does not cross with the pair $(u^i,\upsilon^i)$ $\forall 1 \leq i \leq n$, or with the pairs $(\ell^j, r^j)$ $\forall 1 \leq i \leq m+n$. Hence we can always draw red edges between the pairs $(u^i,\upsilon^i)$, $(\ell^j, r^j)$, and $(s, \ell^1)$. There are $n$ red edges of type $(u^i,\upsilon^i)$, $m+n$ red edges of type $(\ell^j, r^j)$, and a single red edge of type $(s, \ell^1)$. This makes the total number of such red edges $m+2n+1$. \qed

\subsection{Proof of Lemma~\ref{vimp}}
\noindent
In a red blue graph of a perfect block sorting schedule $G(\pi, \mathcal{S}_{perfect})$, all the $m+2n+2$ blue components mentioned in Lemma~\ref{vimp} are connected. Claim A, B, and C of the proof of Lemma $11$ of~\cite{blocksort03} prove:
\begin{enumerate}
\item The union of these $m+2n+2$ blue components form a connected subgraph of $G(\pi, \mathcal{S}_{perfect})$.
\item The only way to connect all of these $m+2n+2$ blue components in $G(\pi, \mathcal{S}_{perfect})$ is to have all the edges of set $\mathbf{E}$. 
\end{enumerate}
The two above implications lead to the fact that $\mathbf{E}$ is the only set of edges which can connect all of these in $m+2n+2$ blue components in any $G(\pi, \mathcal{S})$. Hence any edge absent from $G(\pi, \mathcal{S})$ will imply at least one disconnected blue component in $G(\pi, \mathcal{S})$.
\qed

\subsection{Proof of Lemma~\ref{joinpairspq}}

\begin{enumerate}
\item We have $\pi(p_k^j) < \pi(p_i^j) < \pi(q_k^j) < \pi(q_i^j)$, for $1 \leq i \leq n$, $1 \leq j \leq m$. Hence by Lemma~\ref{red2} they cross.

\item We have $p_i^j < \bar{p}_i^k < q_i^j < \bar{q}_i^k$ for $1 \leq i \leq n$, $1 \leq j \leq m$, and $1 \leq k \leq m$. Hence by Lemma~\ref{red3} they cross. \qed
\end{enumerate}

\section{Omitted Proofs from Section~\ref{sec5}}\label{sec4}

\subsection{Proof of Lemma~\ref{nub}}
\noindent
To prove this, we just observe that for $k=1$ which is the minimum value $k$ can have, {\sc $k$-Block Merging} reduces to {\sc Block Merging}. For the maximum value $k$ can have, which is simply equal to the number of increasing sequences $\mathbb{S}$ has, {\sc $k$-Block Merging} reduces to {\sc Block Sorting}. Hence the above inequalities holds for any $1 \leq k \leq (\#$ of increasing sequences in $\mathbb{S})$. \qed

\subsection{Proof of Lemma~\ref{bm}}
\noindent
The proof sketch of Lemma~\ref{bm} is based on the fact that a block is allowed to be moved in {\sc Block Merging} if it is contained in at most one increasing sequence. This in fact ensures that at most one edge can be added to $c(\mathbb{S}_{\pi})$ by a block merging move. Also, a valid block merging move in which at least one block is merged adds at least one edge to $c(\mathbb{S}_{\pi})$. Hence $c(\mathbb{S}_{\pi})$ can be increased by exactly $1$ by each block merging move.

\noindent
Further, it has been shown in~\cite{blocksort06} that a block merging move to reduce $c(\mathbb{S}_{\pi})$ by $1$ can be always found in polynomial time. Since, $c(\mathbb{M}_{n})=n-1$, this gives a polynomial time exact algorithm for block merging. \qed

\subsection{Proof of Lemma~\ref{nbm}}
Since we allow a block to be moved in {\sc $k$-Block Merging} if it is contained in at most $k$ increasing subsequences, we can add at most $k$ new edges to 
$c(\mathbb{S}_{\pi})$ by a $k$-block merging move. A block fragmented across $k$ increasing sequences in $\mathbb{S}_{\pi}$ will add $k$ edges to $c(\mathbb{S}_{\pi})$ when moved into one increasing subsequence by a single $k$-block merging move. \qed
\newpage
\section{An Application of {\sc Block Sorting} in OCR}

\begin{figure}\begin{center}
\includegraphics[scale=.85]{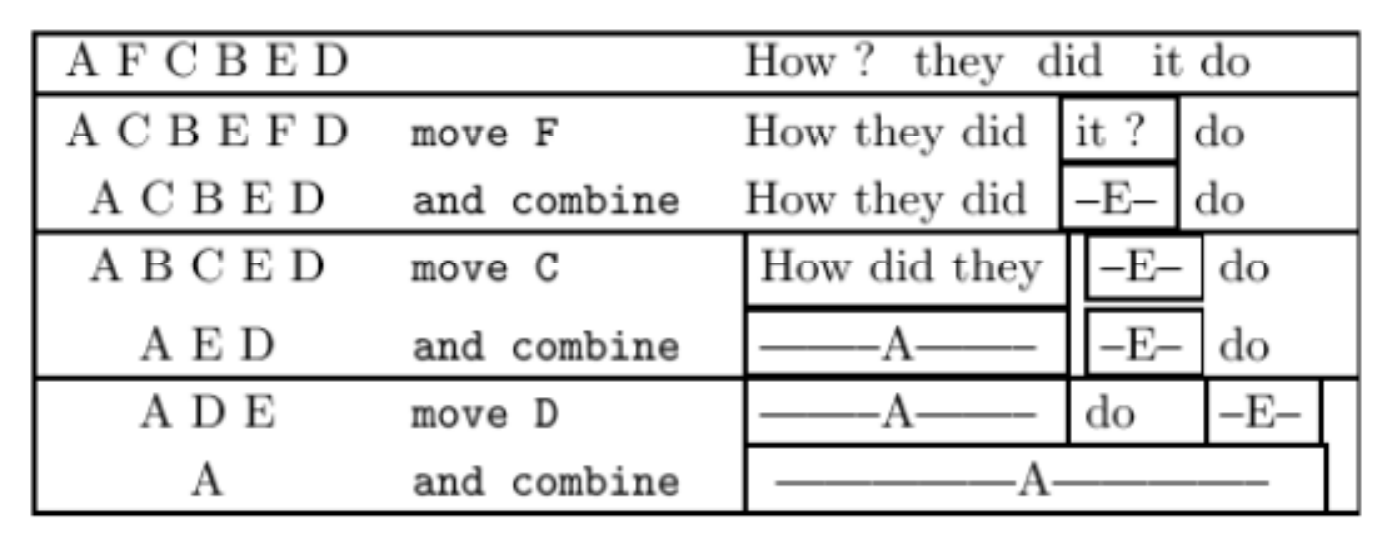}
\caption{\label{ex2} Block sorting string ``How did they do it?" (Reproduced from~\cite{blocksort03}).}
\end{center}
\end{figure}

\noindent
In figure~\ref{ex2}, we illustrate this concept with an example inspired by an application in optical character recognition. Here we have a permutation ``How ? they did it do" recognized, but not in the correct order ``How they did do it ?". We observe that it requires $3$ block moves to sort the permutation by using {\sc Block Sorting}. The blocks are moved and combined with other blocks to form larger blocks at each step. 

\begin{thebibliography}{00}
\bibitem{paper1} R. Gobi, S. Latifi, W.W. Bein. {\em Adaptive Sorting Algorithms for Evaluation of Automatic Zoning Employed in OCR Devices}. In Proceedings of the 2000 International Conference on Imaging Science, Systems, and Technology - CISST 2000, CSREA Press, (2000), 253--259.

\bibitem{blocksort03}  W.W. Bein, L.L. Larmore, S. Latifi, and I.H. Sudborough. {\em Block sorting is hard.} International Journal of Foundations of Computer Science, 14(3):425-437, 2003.

\bibitem{blocksort06} M. Mahajan, R. Rama, V. Raman, and S. Vijaykumar. {\em Approximate Block Sorting.} International Journal of Foundation of Computer Science, 2006: 337-356.   

\bibitem{blocksort05}  W.W. Bein, L.L. Larmore, S. Latifi, and I.H. Sudborough. {\em A Faster and Simpler 2-Approximation Algorithm for Block Sorting.} Lecture Notes in Computer Science 3623, Springer Verlag, 2005, pages 115-124. 

\bibitem{paper2} V. Bafna and P. A. Pevzner. {\em Sorting by Transpositions}. SIAM Journal of Discrete Mathematics. Vol. 11, No. 2, pp. 224--240, May 1998.

\bibitem{paper3} I. Elias, T. Hartman. {\em A 1.375-Approximation Algorithm for Sorting by Transpositions}. IEEE/ACM Transactions on Computational Biology and Bioinformatics, vol. 3, no. 4, pp. 369-379, Oct.-Dec. 2006, doi:10.1109/TCBB.2006.44.

\bibitem{paper4} L. Bulteau, G. Fertin, I. Rusu. {\em Sorting by Transpositions is Difficult.} In Automata, Languages and Programming, Vol. 6755 (2011), pp. 654-665. 

\bibitem{paper5} D. A. Christie. {\em Genome Rearrangement Problems.} PhD Thesis, University of Glasgow, 1999.

\bibitem{paper6} L. S. Heath and J. P. C. Vergara. {\em Sorting by Short Block-Moves}. Algorithmica. Volume 28, Number 3, 323-354, DOI: 10.1007/s004530010041

\bibitem{paper7} H. Jiang and D. Zhu. {\em A (1+e)-Approximation Algorithm for Sorting by Short Block-Moves.} International Joint Conference on Computational Sciences and Optimization, 2009. CSO 2009. 24-26 April 2009, pp. 580--583, Sanya, Hainan. 


\end{thebibliography}
\end{document}